\documentclass[pra,twocolumn,tightenlines,nofootinbib]{revtex4}
\usepackage{bm,dcolumn,amsmath,graphicx}

\newcommand{\eref}[1]{Eq.~(\ref{#1})}

\begin{document}

\title{Comment on ``21-cm Radiation: A New Probe of Variation
in the Fine-Structure Constant''}

\date{ \today }

\begin{abstract}
 Khatri and Wandelt reported that change in the value of $\alpha$
by 1\% changes the mean brightness temperature $T_b$ decrement of the CMB
 due to 21 cm
absorption by 5\% over the redshift range z $<$ 50. A drawback of their
 approach is that the dimensionful parameters are used. Changing of units leads
 to the change of the magnitude and even sign of the effect.
 Similar problems may be identified in a large
number of other  publications which consider limits on the variation of
 $\alpha$ using dimensionful parameters.
 We propose a method to obtain consistent results and provide an estimate
 of the effect.
\end{abstract}

\maketitle
In the Letter~\cite{KhaWan07} Khatri and Wandelt investigated
the effect of variation in the value of the fine-structure constant $\alpha$
at high redshifts and reported that change in the value of $\alpha$
by 1\% changes the mean brightness temperature $T_b$ decrement of the CMB due to 21 cm
absorption by 5\% over the redshift range z $<$ 50.
 The authors made this
 conclusion starting from expressions
\begin{eqnarray}
T_b = \frac{(T_s - T_\gamma) \tau}{1+z} , \qquad
\tau \equiv \frac{3 c^3 \hbar A_{10} n_H}
                 {16 k_B \nu^2 H T_s},
\label{Tb}
\end{eqnarray}
where $T_s$ is the spin temperature, $T_\gamma$ is the radiation temperature
given by $T_\gamma \approx 2.73(1+z)$K,  $z$ is the redshift,
 $c$ is the speed of light,
$k_B$ is the Boltzmann constant, $\nu$ and $A_{10}$ are the frequency and
 the probability of the hyperfine ($1S_{1/2}, F=1$) -- ($1S_{1/2}, F=0$)
 transition in hydrogen;
$n_H$ is the total number density of hydrogen nuclei, and $H$ is the Hubble parameter
at redshift $z$.
Based on the following estimates
$\nu \sim \alpha^4$, $\qquad A_{10} \sim \alpha^{13}$,
and using \eref{Tb} the authors of Ref.~\cite{KhaWan07} come to conclusion
that
 $T_b \sim A_{10}/\nu^2 \sim \alpha^5$
$\label{Tb_al}$
giving $\Delta T_b/T_b$ = 5\% for 1\% change in $\alpha$.

A drawback of this approach is that the dimensionful parameters are used.
Up to a numerical constant, the frequency\\ 
$ \nu \sim \alpha^4 g_I (m/m_p)(m c^2/h)$ ,\\
where $m$ is the electron mass, $m_p$ is the proton mass and $g_I$ is the
proton magnetic $g$ factor. Thus, in units $m c^2$ we have $\nu \sim \alpha^4$.
In the atomic units (which are more natural for the atomic
 problem we deal with) $ \nu \sim \alpha^2 $.
  If the frequency is measured in conventional
SI units  s$^{-1}$ ( defined using the Cs atom hyperfine frequency)
 $\nu \sim \alpha^{-0.83}$ \cite{Tedesco}, i.e. the effect has an opposite
 sign.

 Moreover, even in units of  $m c^2$ used by the authors of
 \cite{KhaWan07}
 we can obtain a different result. The temperature $T_b$ is actually
 defined using  the observed intensity
 $I_\nu = 2 k_B \nu^2 T_b/c^2  \sim \alpha^{13}$. Therefore,
 $\Delta I_\nu / I_\nu $ = 13\% for 1\% change in $\alpha$.

 Similar problems may be identified in a large
number of other  publications which consider limits on the variation of
 $\alpha$ using dimensionful parameters. Simple replacement of the electric
 charge squared by $\alpha$ in all equations and its variation gives
 meaningless results. 

Such problems do not appear in the laboratory (atomic clocks) and quasar
absorption spectra measurements of  the variation of the fundamental constants.
From the very beginning these studies deal with the dimensionless ratios
 of the atomic transition frequencies \cite{Tedesco,Dzuba,Webb}.
 The atomic unit of energy $m e^4/\hbar ^2$
 cancels out in the  ratios of the frequencies.
 The dependence on $\alpha=e^2/\hbar c$
appears from the dimensionless ratio of the relativistic corrections
 to the atomic unit of energy.

  In the quasar spectra
 analysis \cite{Webb}  many frequencies are used to find $\alpha$ variation
 and redshift. The redshift cancels out in the ratios of the frequencies.
 Therefore, $\alpha$ variation is determined by these ratios.
 
 Let us try a similar approach of using dimensionless ratios for the variation 
of the brightness temperature. It is reasonable to start from
the ratio $T_b/T_\gamma$ which is given by (see \eref{Tb})
\begin{eqnarray}
X_T \equiv \frac{T_b}{T_\gamma} = \frac{(1 - T_\gamma/T_s)}{1+z} \,
\frac{3 c^3 \hbar A_{10} n_H}{16 k_B \nu^2 H T_\gamma}.
\label{TbTgam}
\end{eqnarray}
The dimensionless atomic parameter here is\\
 $X_A \equiv A_{10}/\nu \sim \alpha^9 g_I^2 (m/m_p)^2$.\\
Fortunately,  Eq. (1) from \cite{KhaWan07}
 which determines the ratio
$T_\gamma/T_s$, depends on the same atomic parameter $X_A$.
 
The  remaining dimensionless parameter\\
$X_H=(c^3 \hbar n_H)/ (k_B \nu H T_\gamma)$\\
 contains the hydrogen density $n_H=\eta n_\gamma \sim \eta T_\gamma ^3 $
 where $n_\gamma$ is the photon density.
 The numerical value of the
 proton-to-photon number ratio  $\eta$ has been obtained from CMB (or BBN)
data in assumption that there was no variation of the fundamental constants.
 If there was any variation, this value would be different (for example,
the equation for the ionization equilibrium contains combination 
$\eta \alpha^3$). Therefore, $X_H$ is actually sensitive to the value of
$\alpha$ at CMB era where $\delta \alpha$ could be larger. We leave this
 complicated problem for a future study. Note that in the case of BBN
 we calculated how the best fit of $\eta$ is affected by the change
of the fundamental constants \cite{Dmitriev}.
 Now it seems reasonable to
 assume that the  dependence
 of $X_H$ on the fundamental
constants is weaker than  the dependence of $X_A$
which is enhanced by an order of magnitude by  the factor $\alpha^9$.
 Then we obtain\\
$\delta X_T/X_T \sim 9 \delta X_{\alpha}/ X_{\alpha}$,\\
where $X_{\alpha}=\alpha [g_I (m/m_p)]^{2/9}$. The dependence of the proton
g-factor and $m/m_p$ on the fundamental constants has been presented in Ref.
\cite{Tedesco}: 
 $g_I \sim (m_q/\Lambda_{QCD})^{-0.1}$, and
 $m/m_p  \sim (m /\Lambda_{QCD}) (m_q/\Lambda_{QCD})^{-0.05}$ where $m_q$
 is the quark mass and $\Lambda_{QCD}$ is the strong interaction QCD scale.

A rough estimate of the redshift dependence of the $X_{\alpha} $ variation
effect  may be extracted from  the $z$-dependence
of the $\alpha$ variation effect presented on Fig. 2 of Ref. \cite{KhaWan07}.
  Indeed, according to the estimates given above the relative difference
 in these effects is $\sim 9/5$.

In principle, one may try to reduce the problem to that calculated
 in Ref. \cite{KhaWan07} by considering (instead of $X_A$) a different
 dimensionless parameter\\
 $X_B \equiv A_{10} T_\gamma/\nu^2 \sim \alpha^5 g_I (T_\gamma/m_p c^2)$.\\
A minor problem here is that the equation for the spin temperature  $T_s$
presented  in Ref. \cite{KhaWan07} contains $X_A$ (rather than  $X_B$).
 A major problem is how to reduce the variation of the
ratio $T_\gamma/m_p c^2$ to the variation of the dimensionless fundamental
 constants. The CMB temperature is the red-shift-dependent phenomenological
 parameter which depends on units we use, and these units may be
 time-dependent. Even if the variation of $X_B$ is ``detected'',
 it would be  hard to provide an  interpretation in terms
of theories of the fundamental constants  variation.

V.V. Flambaum, S.G. Porsev\\
School of Physics, University of New South Wales,
Sydney 2052, Australia.

\begin{thebibliography}{2}
\expandafter\ifx\csname natexlab\endcsname\relax\def\natexlab#1{#1}\fi
\expandafter\ifx\csname bibnamefont\endcsname\relax
  \def\bibnamefont#1{#1}\fi
\expandafter\ifx\csname bibfnamefont\endcsname\relax
  \def\bibfnamefont#1{#1}\fi
\expandafter\ifx\csname citenamefont\endcsname\relax
  \def\citenamefont#1{#1}\fi
\expandafter\ifx\csname url\endcsname\relax
  \def\url#1{\texttt{#1}}\fi
\expandafter\ifx\csname urlprefix\endcsname\relax\def\urlprefix{URL }\fi
\providecommand{\bibinfo}[2]{#2}
\providecommand{\eprint}[2][]{\url{#2}}

\bibitem[{\citenamefont{Khatri and Wandelt}(2007)}]{KhaWan07}
\bibinfo{author}{\bibfnamefont{R.}~\bibnamefont{Khatri}} \bibnamefont{and}
  \bibinfo{author}{\bibfnamefont{B.~D.} \bibnamefont{Wandelt}},
  \bibinfo{journal}{Phys. Rev. Lett.} \textbf{\bibinfo{volume}{98}},
  \bibinfo{pages}{111301} (\bibinfo{year}{2007}).

\bibitem{Tedesco} V.V. Flambaum and A.F. Tedesco, Phys.Rev.
 C {\bf 73}, 055501 (2006). 
\bibitem{Dzuba} V. A. Dzuba,  V.V. Flambaum, J. K. Webb ,
 Phys. Rev. Lett. {\bf 82}, 888, 1999;  Phys. Rev. A59, 230, 1999.

\bibitem{Webb}
  J. K. Webb , V.V. Flambaum, C.W. Churchill, M.J. Drinkwater, and J.D. Barrow,
 Phys. Rev. Lett. {\bf 82}, 884, 1999.
\bibitem{Dmitriev} V.F. Dmitriev, V.V. Flambaum, J.K. Webb. Phys. Rev.
 D {\bf 69}, 063506 (2009).
\end{thebibliography}

\end{document}